## Influence of ligand field and correlation on the electronic structure of NiO and CoO from DFT+DMFT calculations

Daniel Mutter[1], Frank Lechermann[2], Daniel F. Urban[1,3], and Christian Elsässer[1,3]

[1]Fraunhofer IWM, Wöhlerstraße 11, D-79108 Freiburg, Germany

[2]Theoretische Physik III, Ruhr-Universität Bochum, D-44780 Bochum, Germany

[3]Freiburg Materials Research Center (FMF), University of Freiburg, Stefan-Meier-Straße 21, D-79104 Freiburg, Germany

## Abstract

The intriguing physics and rich application potential of strongly correlated first-row transition metal oxide compounds result from the complex interplay of several factors that influence the electronic structure. To shed light on the effect of composition, structure, and correlation strength, we apply a well-established charge self-consistent combination of density functional theory and dynamical mean field theory, which has proven to give electron binding energies in good agreement to experimentally derived excitation spectra. For paramagnetic NiO and CoO, we analyze the effect of rock-salt and zincblende structures and their different ligand fields on the spectral functions. By varying the value of the interaction parameter $U$, different correlation strengths among the transition-metal $3d$ electrons are considered, as well as the effect of additionally accounting for correlations in the oxygen $2p$ orbitals by a self-interaction-correction pseudopotential scheme. It is shown that this combined DFT+sicDMFT approach leads to spectral functions and band gaps close to experimental data, thereby offering a time-efficient alternative to using a fully correlated model which would treat both, metal and oxygen ligand electrons within DFT+DMFT.

## 1. Introduction

Transition metal oxide compounds (TMOs) have attracted tremendous attention since decades due to the fascinating physical phenomena, such as the Mott metal-insulator transition, multiferroicity, or spin-orbital superexchange [1], [2], [3], [4]. Considering specifically NiO and CoO, potential applications include thermoelectricity [5], nanostructured optoelectronics [6], electrochemical energy storage devices [7], and catalysts for the splitting of water [8], [9], [10] or higher alcohols [11]. The key to this vast space of research and application lies in the electronic structure of the TMOs. For example, Hong et al. [12] showed that for perovskite-type TMO compounds, the current density of the oxygen evolution reaction taking place at the surfaces is directly related to an electronic bulk property, namely the charge-transfer energy between occupied O($2p$) and empty TM($3d$) electronic states. The charge-transfer energy can therefore serve as a descriptor for the material's catalytic activity.

In general, in the narrow bands resulting from the partially filled $3d$ shells of the TM ions in TMOs, the electron's tendency to delocalize, best described in a mean-field notion of non-interacting particles, competes with correlation effects, i.e., with direct influences on a moving electron by the presence of other electrons [13]. In addition, the oxygen ions surrounding the TM ions create a ligand field in which the TM-$d$ orbitals split into submanifolds of different energies. The hierarchy of those energies depends on the crystal structure [1]. TM($3d$) and O($2p$) orbitals are hybridized to an extent that depends on the position of the TM element in the periodic table, or, accordingly, on the degree of filling of the $d$ orbitals with electrons and the strength of the Coulomb attraction of the ionic core [14].

The relative binding energies of O($2p$)- and TM($3d$)-like electrons define whether the insulating behavior of the TMO is of Mott-Hubbard or charge-transfer type [15]. In the Mott-Hubbard case, the band gap originates from a transition between a ground state with two neighboring TM atoms $d^n d^n$ and an excited state $d^{n-1}d^{n+1}$. Those states are separated in energy according to the electron-electron interaction (correlation) strength $U$ if the process creates an empty and a doubly occupied orbital from two singly occupied orbitals (corresponding to the lower and upper Hubbard bands, respectively). In the charge-transfer case, the energy difference $\Delta$ between a final $d^{n+1}$ state and the O($2p$) state is smaller than $U$, i.e., the excitation defining the band gap corresponds to the creation of a ligand hole $L$ and an occupied $d$-state. While "early" TMOs (such as VO) are Mott-Hubbard insulators, "late" TMOs (such as NiO) exhibit charge-transfer behavior [16], [17].

Binding energies of excited electronic states in TMOs are defined by the intricate interplay of several effects, namely quasiparticle band formation, correlation strength, ligand field, crystal structure, hybridization, and magnetism. Experimentally, binding energies can be measured by photoemission spectroscopy (UPS/XPS) and by bremsstrahlung isochromate spectroscopy (BIS), probing occupied and unoccupied states, respectively [14], [18], [19]. Electronic-structure calculations based on density functional theory (DFT) with exchange-correlation treated by the local density or generalized gradient approximations (LDA or GGA) fail in reproducing the experimental spectra of strongly correlated TMOs. In many cases, such calculations predict metallic instead of insulating ground states [20]. A considerable improvement was achieved by combining DFT in a charge self-consistent way with the dynamical mean field theory (DMFT) [21], [22]. DMFT treats the electron-electron interaction explicitly within a subspace of correlated orbitals, which accounts for dynamical charge and spin fluctuations. Especially for paramagnetic (high-temperature) configurations, DFT+DMFT can well reproduce the details of measured excitation spectra and band gap values of TMOs [23], [24], [25], considerably better than a static mean-field approach to correlation as in DFT with LDA and GGA, or in DFT+U [26], [27], [28].

DMFT captures the simultaneous localized and delocalized character of the correlated electrons by mapping the electronic problem onto a lattice model in which electrons can hop between an impurity site and a surrounding bath. [29]. Both TM($3d$) and O($2p$) orbitals of a TMO can in general be treated in this way in a fully correlated model. However, due to the high computational effort in solving the quantum impurity problem, DMFT is typically only applied to the subspace of the TM-$d$ states, where the correlation effects are most relevant. In our approach, we additionally consider correlation effects in the O($2p$)- and in the hybridized $pd$-orbitals effectively by using a self-interaction corrected (SIC) pseudopotential [30], [31] for oxygen. As it treats oxygen correlation on the pseudopotential level, this *DFT+sicDMFT* approach is computationally considerably more efficient than a fully correlated DMFT model including TM($3d$) and O($2p$) electrons. For paramagnetic NiO, DFT+sicDMFT was demonstrated previously by authors of this work to improve the description of the excitation spectra with respect to experiments, especially concerning the size of the band gap [32]. The DFT+sicDMFT method was further successfully applied to a series of nickelates and cuprates [33], [34], [35]. In a recent work, Carta et al. [36] also describe the importance of considering ligand on-site interactions in DFT+DMFT calculations for obtaining charge-transfer energies and insulating states to get good agreement with measured spectra. Instead of SIC, they applied a Hartree-Fock-like correction to the O($2p$) orbitals and demonstrated their approach for perovskite TMOs.

The purpose of the present paper is to analyze different influence factors on the spectral functions of excited electronic states for the TMOs NiO and CoO by performing DFT+sicDMFT calculations. Both compounds are considered in their paramagnetic states in the cubic rock-salt (RS) and zincblende (ZB) structures. In those structures, TM($3d$) ions feel octahedral (RS) and tetrahedral (ZB) ligand fields of oxygen ions. We further analyze the effect of SIC and of the strength of the electron-electron interaction parameter $U$ on the TM($3d$) orbitals. The paper is organized as follows: in Section 2, we give a descriptive overview and provide computational details of the applied DFT+sicDMFT method. Section 3 presents and discusses results of spectral functions: in 3.1 in more detail for RS-CoO and in 3.2 comparatively for both (CoO and NiO) compounds with both (RS and ZB) structures. In 3.3, characteristic features of all calculated spectral functions for the different correlation-related settings are compared. A summary is given in Section 4.

## 2. Methods and Models

The DFT+DMFT approach reveals low and high energy excitations as well as quasi-particle features in the spectral functions of TMOs with strongly correlated electrons. A central quantity of DMFT is the self-consistently determined, energy-dependent dynamical mean field $\mathcal{G}_0$, which captures dynamical correlations in the form of charge and spin fluctuations. For a detailed description of the quantum impurity problem and of the formalism to combine DMFT with DFT, the reader is referred to the extensive literature on those topics [29], [37], [38], [39]. The methods applied in this work are mainly those described in Refs. [22], [40]; they are only briefly summarized in the following.

From a technical point of view, coupling DFT with DMFT needs the objects appearing in the equations of both approaches (wavefunctions, operators, Green's functions) to be expressed by a common basis. While DFT is typically formulated in the Kohn-Sham basis of delocalized Bloch waves, a basis of localized orbitals is more appropriate for DMFT owing to the origin of strong correlation effects in spatially confined bands. A connection between the two descriptions can be made by considering correlation effects only within a subspace of the full Hilbert space of the electronic system and by decomposing the correlated orbitals spanning this subspace in the Kohn-Sham basis. The number of Kohn-Sham basis functions needs to be restricted to make the calculations feasible. As an example, Figure 1 shows for CoO in the RS structure the band structure together with the site- and orbital-projected densities of states obtained from an LDA-DFT calculation. It was calculated with the mixed basis pseudopotential code [41], [42], which we employed throughout this work also for the DFT part of the DFT+DMFT calculations.

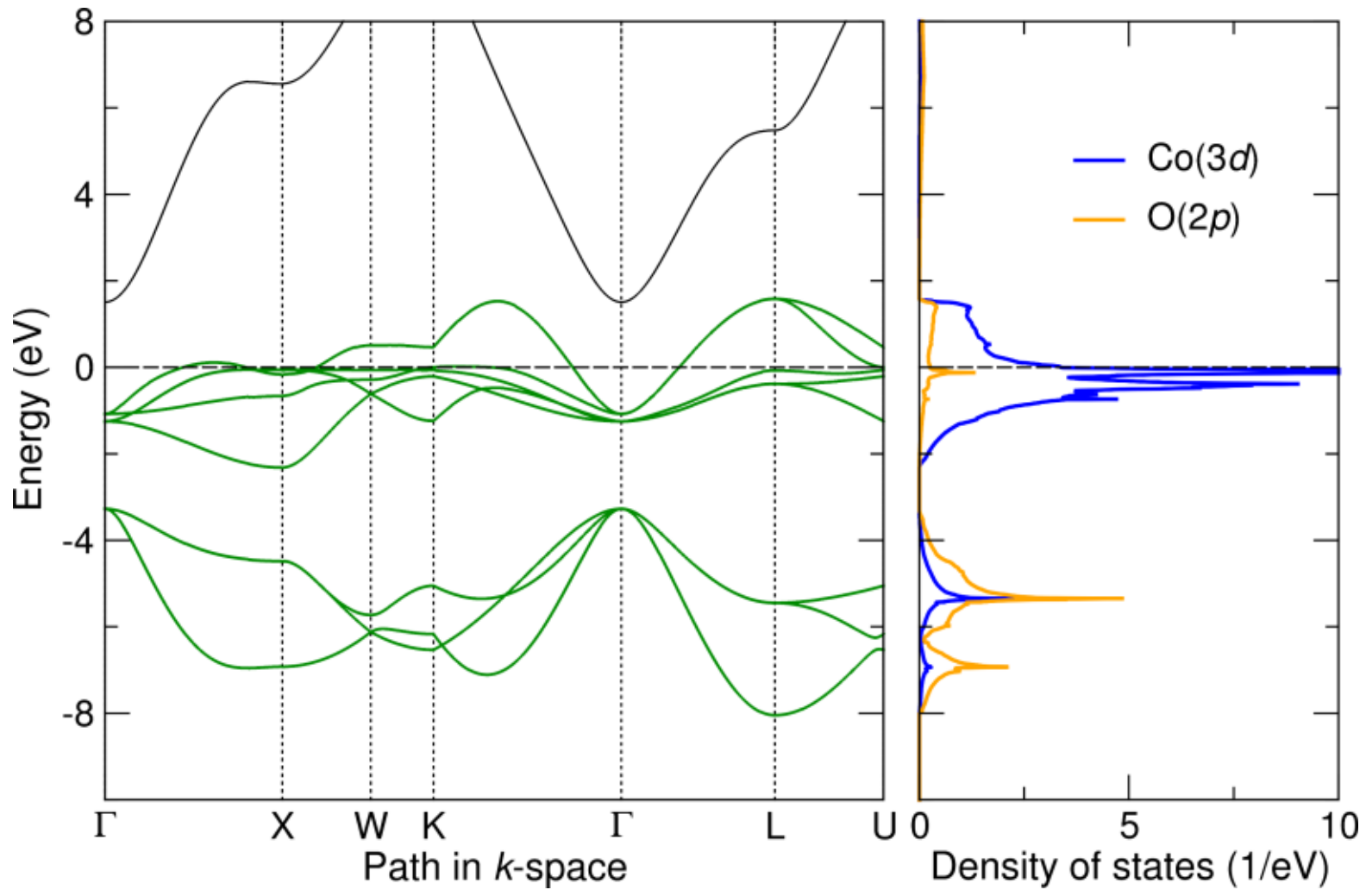

*Figure 1. Electron band structure (left) and site- and orbital-projected densities of states (right) of CoO in the RS structure, calculated by LDA-DFT. Energies are given with respect to the Fermi level.*

Since correlations in a TMO compound are most relevant between the TM($3d$) electrons, it is reasonable to choose the corresponding atomic-like orbitals being centered around the TM and having $d_{xy}$, $d_{xz}$, $d_{yz}$, $d_{x^2-y^2}$, and $d_{z^2}$ angular momentum character for constructing the correlated subspace. In the example shown in Figure 1, the corresponding bands are located in the energy range between $-2$ eV and 1 eV with respect to the Fermi level, but there is also a non-negligible $d$-orbital weight in the energy range of the O($2p$) states, from $-8$ to $-4$ eV, due to hybridization. One therefore needs to include at least the Bloch states belonging to those 8 bands [3 x O($2p$) and 5 x Co($3d$) in the primitive cell] to appropriately describe the correlated $3d$-orbitals in the Kohn-Sham basis. Atomic-like orbitals represented in a truncated Kohn-Sham basis are an alternative to the widely used maximally localized Wannier functions for defining the correlated subspace in DFT+DMFT calculations [22], [43], [44], [45]. We applied the former approach for the TMOs studied in this work.

Within the so-defined correlated subspace, we utilize the rotationally invariant Slater Hamiltonian for the generalized multi-orbital Hubbard problem [46]. Assuming nearly spherical symmetry, the matrix elements are formulated via parametrized Slater integrals $F^k$. Setting $F^4/F^2 = 0.625$, one is left with two parameters to be chosen, namely the Hubbard parameter $U$ and Hund's exchange coupling parameter $J$ [47], [48]. $U$ values applied in DMFT studies of NiO range between 7 eV [25] and 10 eV [49], and $J$ values are typically chosen around 1 eV (see Table I in Ref. [50] for an extensive list). Several authors report interaction parameters calculated from first principles based on the constrained random phase approximation [20], [50], [51]. For a fully correlated model of Ni($3d$) and O($2p$) electrons (named $dp$-$dp$, or just $dp$ model in those works), they obtain $U$ values close to 10 eV for NiO and about 0.5 eV lower values for CoO.

In this work, in addition to treating dynamic correlation between the electrons in the TM($3d$) orbitals with DMFT, we apply the SIC scheme to account for correlation effects in the O($2p$) states, which is motivated by the strong hybridization between TM($3d$) and O($2p$) orbitals in TMOs. SIC serves as another means to introduce an on-site repulsion term $U_{pp}$ to the O($2p$) orbitals as in DFT+U, with the same effect of impeding the respective electrons to delocalize too strongly. With DFT+U being based on the Hubbard model, which explicitly introduces the repulsive U term as an effect of electron-electron correlation, we also consider SIC as an alternative scheme to treat correlation effects. Within SIC, the angular momentum ($l$) specific pseudopotentials $V_l$ of the atoms of choice are adjusted by subtracting, to some extent (specified by a parameter $w_l$) the Hartree and exchange-correlation potentials $V_{\mathrm{H}}[n_l]$ and $V_{\mathrm{xc}}[n_l]$ for the electron density $n_l$ corresponding to the specified atomic orbital [30]. This procedure strongly diminishes the spurious interaction of electrons with themselves present in LDA/GGA-based DFT. Since the basic SIC formalism is developed for free atoms, further refinement is necessary to adapt it to atoms on sites in a crystal structure. This is done by an additional dimensionless factor $\alpha$ to account for the strength of the solid-state adjustment for a specific element and compound [31]. In our DFT+sicDMFT calculations, we set $\alpha = 0.8$, $w_{l=s}(0) = 1.0$ and $w_{l=p}(0) = 0.8$. This corresponds to the O($2s$) states being fully corrected (100 %) and the O($2p$) states by 80 %. Improved band gaps with respect to experimental values have been obtained with this choice of parameters for SIC-DFT calculations of ZnO [31], $TiO_2$ [52] , Zn-Sn-O systems [53], and $Zn_{1-x}Mg_xO$ compounds [54]. With $Ti^{4+}$ ($d^0$) in $TiO_2$ and $Zn^{2+}$ ($d^{10}$) in ZnO marking the two ends of filling the $3d$ shell with electrons, the choice of SIC parameters for O has proven to be highly transferable in the above cited SIC-DFT calculations of TMOs. This also holds for different structures, as shown in Ref. [52], where $TiO_2$ was studied in anatase and rutile configuration with the same SIC settings for O, and band gaps were obtained with less than 10% deviation with respect to experimental values. Note that the SIC parameters for the $d$-electrons of the TM elements change between $w_{l=d} = 0$ for Ti [52] and $w_{l=d} = 1$ for Zn [31]. Instead of this full SIC treatment of static correlations we correct only the O($2p$) pseudopotentials with SIC and treat the correlation in the TM($3d$) orbitals by DMFT. The above-mentioned SIC settings also lead to spectral functions of paramagnetic NiO obtained in a DFT+sicDMFT calculation that agree well with experimental spectra [32].

Having defined the correlated subspace and the interaction settings for the systems of choice, the equations underlying the charge self-consistent DFT+DMFT method can be solved iteratively. In the approach of this work, the dynamical mean field, self-energy, and Green's function of the DMFT impurity problem are calculated with the continuous-time Quantum Monte Carlo solver CTHYB, as implemented in the TRIQS package [55]. Then, the LDA-DFT step is performed with the mixed basis pseudopotential code using an updated charge density. This results in a new Kohn-Sham potential and basis functions as input for the next DMFT step [22]. Double counting of correlation contributions from the LDA exchange-correlation potential and DMFT was corrected in the fully localized limit [24], [47] which is also widely applied in DFT+U. After pre-convergence runs with lower numbers of Quantum Monte Carlo steps, the number of steps was increased to values between $5 \cdot 10^8$ and $2 \cdot 10^{10}$ for reaching converged results with respect to chemical potential, self-energy, orbital occupation, and charge density with high accuracy.

In DMFT, Green's functions and self-energies are typically formulated in terms of discrete imaginary Matsubara frequencies ($i\omega_n$), which allows to study temperature-dependent correlation effects. We derived real-frequency dependent spectral functions $A_\nu(\omega)$ and $A_m^{\mathrm{imp}}(\omega)$ by analytical continuation using the maximum entropy method [56] to the final band- and orbital-projected Green's functions $G_{\nu\nu}^{\mathrm{Bl}}(\boldsymbol{k}, i\omega_n)$ and $G_{mm}^{\mathrm{imp}}(i\omega_n)$, respectively, of the converged DFT+DMFT calculation. Here, "Bl" stands for Bloch, "imp" for impurity, $\boldsymbol{k}$ is a vector in reciprocal space, and $\nu$ denotes the index of the band from the truncated number of active bands used to decompose the localized orbitals $m$ in the Bloch basis. The *local* spectral function $A_m^{\mathrm{imp}}(\omega)$ can be identified as the DMFT analog of the partial density of states of DFT. However, experimental excitation spectra are better described by the *total* spectral function $A(\omega)$, which includes the contributions from all bands $\nu$ via $A(\omega) = \sum_\nu A_\nu(\omega)$ [40]. Using normalized $\boldsymbol{k}$-integrated projection numbers $a_\nu^m$ of the localized TM($3d$) orbitals $m$ on the bands $\nu$, one can also formulate the contribution of orbital $m$ to the total spectral function $A(\omega)$ as a weighted sum: $A^m(\omega) = \sum_\nu a_\nu^m A_\nu(\omega)$, which leads to a "TM-projected" spectral function $A_{\mathrm{TM}}(\omega) = \sum_m A^m(\omega)$. Analogously, projecting the final Bloch states of the DFT+DMFT calculation on orbitals with O($2p$)-character allows for the formulation of an "O-projected" spectral function.

In this work, we study total, projected, and local spectral functions of NiO and CoO in RS (space group $Fm\bar{3}m$) and ZB structures ($F\bar{4}3m$). Both structures were set up in their primitive (face-centered cubic) unit cells containing 2 atoms, as shown in Figure 2. The lattice parameters were set to $a$(RS-NiO) = 4.17 Å, $a$(RS-CoO) = 4.24 Å, and $a$(ZB-CoO) = 4.55 Å according to experimental values [57], [58]. Note that ZB-NiO is not a stable structure, and it is studied in this work as a model system for a systematic comparison. We set $a$(ZB-NiO) = 4.47 Å to keep the same ratio between ZB and RS volumes as in CoO.

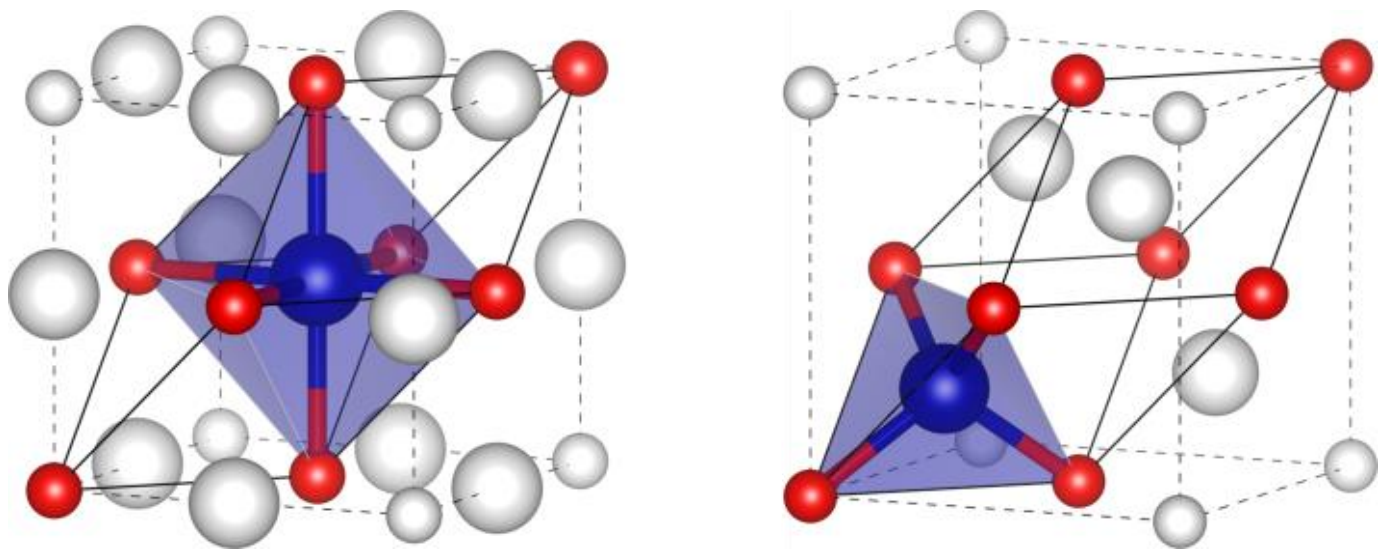

*Figure 2. Conventional cubic (dashed lines) and primitive face-centered cubic cells (solid black lines) of the TMOs considered in this work in RS (left) and ZB structures (right). Large spheres represent the TM (Ni or Co) atoms, and small spheres the O atoms. Atoms colored blue (TM) and red (O) belong to the primitive cells. The octahedral and tetrahedral coordinations of the TM atoms by O atoms are highlighted.*

In the DFT calculations, Brillouin zone integrations were performed for all systems on a 13x13x13 grid constructed by the scheme of Monkhorst and Pack [59]. We used norm-conserving pseudopotentials to describe the interaction of the ionic cores with the valence electrons in the $3d$- and $4s$-shells of Ni and Co, and in the $2s$- and $2p$-shells of O. The mixed basis for expressing those valence (Bloch) states utilized in the mixed basis pseudopotential code consisted of plane waves up to an energy cutoff of 218 eV and localized atomic orbitals of $s$-, $p$-, and $d$-character (the latter of which also served as starting points for constructing the correlated subspace as described above; see also the Appendix in Ref. [40]).

## 3. Results and discussion

### 3.1. Spectral functions of RS-CoO

For RS-NiO, results of DFT+sicDMFT calculations using the method outlined above were already presented and discussed in our previous paper [32]. There we employed $U$ = 10 eV and $J$ = 1 eV to set the matrix elements of the interaction operator between the Ni($3d$) electrons. Moving from right to left in the periodic table of elements, $U$ is expected to decrease due to weaker core attraction and therefore lower localization of the $3d$ electrons. Accordingly, we reduced $U$ heuristically to 9 eV for Co in CoO and left $J$ unchanged. This is consistent with the results of calculations of $U$ and $J$ based on the constrained random-phase approximation [20]. To check the validity of this choice, we performed additional calculations for CoO with a set of lower values, namely $U$ = 6 eV and $J$ = 0.9 eV. Furthermore, we analyzed the effect of SIC on the O($2p$) electrons. Figure 3 shows the total and element-projected spectral functions for CoO in the RS structure obtained by DFT+DMFT and DFT+sicDMFT for an electronic temperature of 580 K. At this temperature, CoO is stable in the cubic, paramagnetic phase [61].

For energies below the Fermi level, the curve shapes of the calculated total spectral functions result from excitation processes corresponding mainly to Co($3d$) and O($2p$) electron emission. Independent of the applied computational settings, the first major peak below the valence band edge has a large Co($3d$) weight, but also a small O($2p$) contribution. This peak is widely ascribed to a $d^7L$ final state, where the emission of a Co($3d$) electron is subsequently screened by the creation of a ligand hole $L$ [62], [63], [64]. The $d^7L$ state can also be reached by emitting an electron from a $d^8L$ state, the latter making up for about 10-20 % of the ground state of RS-CoO [65].

In our calculations, the position of this low energy peak shifts from $-1$ eV to $-1.6$ eV upon increasing $U$ from 6 to 9 eV [Figure 3(a) and (b)]. Leonov et al. [66] report $-1$ eV for a DFT+DMFT calculation with $U$ = 8 eV, but performed at a much higher temperature of 1160 K.

Note that, over the complete energy range, their spectral function is shifted by on average $+0.5$ eV with respect to ours obtained with $U$ = 9 eV at $T$ = 580 K. Applying SIC to the O($2p$) electrons moves the $d^7L$ peak in the opposite direction as the increase of $U$, namely to $-0.6$ eV. An XPS study of CoO [black points in Figure 3(d)] locates the first peak below the valence band edge at $-1.9$ eV [19], whereas the use of lower energy photons in a UPS experiment results in about $-0.7$ eV [18] [Figure 3(d)].

In the latter case, the energy of the incident photons is still above, but close to the $3p \rightarrow 3d$ absorption edge, which triggers an additional electron emission channel via an Auger process. This interferes anti-resonantly with the $d^7L$ final state, resulting in a decrease of the corresponding peak height [18]. As it is also described for NiO, the broad O($2p$) band bunches out more clearly in the UPS spectrum due to that effect [67], with a maximum at $-3.2$ eV ($-4$ eV in the XPS data) for CoO. Our calculations with $U$ = 9 eV and O($2p$)-SIC yield $-3.1$ eV and a shape of the O($2p$) peak resembling the UPS data.

A well-understood signature of strong $d$-$d$ correlation is a high-energy feature often referred to as "satellite" peak, or lower Hubbard band. In the UPS spectrum of RS-CoO, it is located at $-9.5$ eV. The peak corresponds to a final state $d^6$, i.e. a state where one electron is removed from the state $d^7$, the main contribution to the ground state of RS-CoO [62]. In contrast to the $d^7L$ peak, the satellite is resonantly *enhanced* for photon energies close to the $3p \rightarrow 3d$ absorption edge. The spectral function presented by Leonov et al. [66] clearly reveals the lower Hubbard band at an energy close to the experimental values. In our calculation results, it is not visible in the total spectral function, but only very slightly in the Co-projected one derived for $U$ = 9 eV. The application of SIC to O($2p$) states shifts the peak maximum to $-7$ eV, i.e. away from the experimental value. There is, however, also a slight bump visible in the XPS spectrum at $-7$ eV.

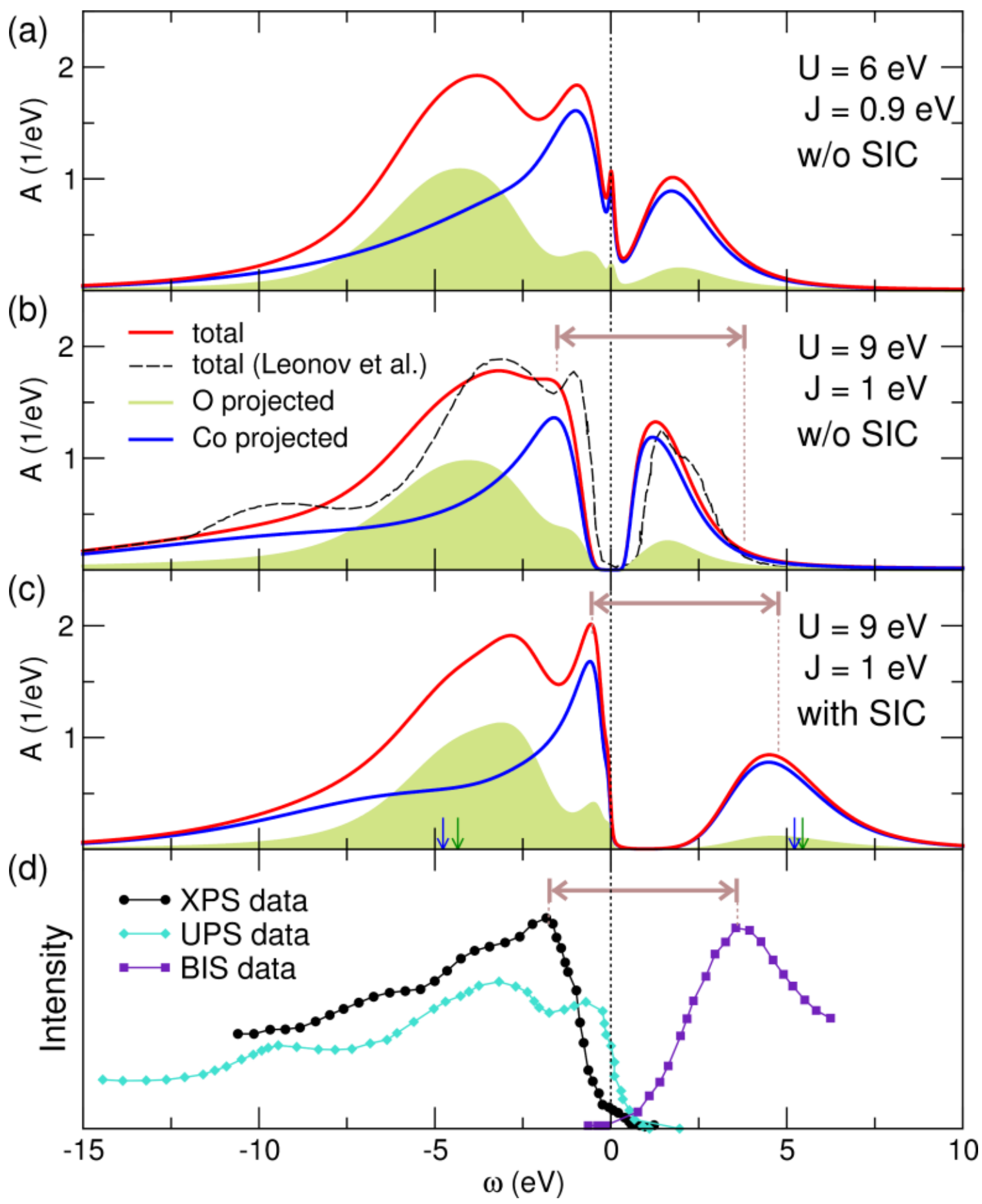

*Figure 3. (a)–(c) Total and element-projected spectral functions of paramagnetic CoO in the RS structure derived from DFT+DMFT or DFT+sicDMFT at $T$ = 580 K as functions of the energy with respect to the Fermi level. (a) $U$ = 6 eV, $J$ = 0.9 eV for the Co(3$d$) electrons without ("w/o") considering SIC on the O(2$p$) electrons; (b) $U$ = 9 eV, $J$ = 1 eV, without SIC. For comparison, the dashed line shows data from Leonov et al.* [66] *for RS-CoO, derived at $T$ = 1160 K for $U$ = 8 eV and $J$ = 0.9 eV and without SIC; (c) $U$ = 9 eV, $J$ = 1 eV, and with O(2$p$)-SIC. The legend in (b) is common to (a), (b), and (c). (d) Experimental spectra (in arbitrary units): XPS (at a photon energy of 1253.6 eV) and BIS (1486.6 eV) from Ref.* [19], *UPS (80 eV) from Ref.* [18] *with lines as a guide for the eye. The horizontal light brown line with arrow tips indicates the distance between the maxima of the XPS and BIS peaks (5.3 eV). It is also shown with the same length in (b) and (c) for comparison. Blue (green) vertical arrows in (c) are placed at the centers of gravity of the occupied [$\omega < 0$] and unoccupied [$\omega > 0$] Co-projected (O-projected) spectral functions.*

To reveal the unoccupied states, an electron can be added to the system in a BIS experiment, resulting in a spectrum which is related to the DMFT spectral function for energies above the Fermi level. The first peak corresponds to a $d^8$ final state. With the O(2$p$) spectral weight being located between the lower and the upper Hubbard band, CoO can be classified as a charge transfer insulator [68] with the charge transfer process $d^7 \rightarrow d^8 L$.

A band gap is visible in the combined XPS and BIS spectrum provided by van Elp et al. [19] [Figure 3(d)], but its size cannot be defined unambiguously. The measured distance between the peak maxima below and above the Fermi level in that work is 5.3 eV. Experimental spectra of Kurmaev et al. exhibit about 6 eV [69]. From our calculation results, it is apparent that a band gap only opens for a large enough value of the Co(3$d$) electron interaction strength $U$ [Figure 3(a) and (b)]. This is mainly due to a reduction of Co(3$d$) spectral weight in the vicinity of the Fermi level. A considerable shift of the unoccupied peak to a higher value is apparent when additionally considering the O(2$p$)-SIC, with the difference to the first occupied main peak of 5.1 eV.

Lechermann *et al.* [32] described a similar gap-opening effect in DFT+sicDMFT calculations of RS-NiO. As explained there, the DFT+DMFT formalism attempts to lower the energy of the electronic system by increasing (virtual) charge fluctuations between occupied O($2p$) and unoccupied Ni($3d$)-$e_g$ states (upper Hubbard band). This effect reduces the distance between those states. Of course, the $d$-$d$-interaction within the $e_g$-orbitals ensures a finite gap, which however turns out to be smaller as the gap observed in experimental spectra. It is assumed that this discrepancy originates from too strongly delocalized O($2p$) electrons that support the charge fluctuations between Ni($3d$)-$e_g$ and O($2p$) states. The application of SIC on the latter reduces this delocalization and hence impedes too strong charge fluctuations, which can lower the gap, by correcting the DFT-inherent erroneous self-interaction between the electrons. As a result, the DFT+sicDMFT approach leads to larger band gaps compared to standard DFT+DMFT.

Accordingly, the value of 5.1 eV obtained in our calculations is close to the experimental range of 5 – 6 eV. For comparison, for that difference between the occupied and unoccupied peaks of lowest energy, previous DFT+DMFT studies of RS-CoO without O($2p$)-SIC reported considerably lower values of about 2.5 eV applying $U$ = 8 eV and $J$ = 0.9 eV [66] or $U$ = 6 eV and $J$ = 1 eV [70], 3 eV ($U$ = 8 eV, $J$ = 0.9 eV) [71], and 3.8 eV ($U$ = 8 eV, $J$ = 1 eV) [24].

### 3.2. Influence of structure and composition

Having analyzed the electronic structure of RS-CoO in detail in the previous section, we now compare the spectral functions of CoO and NiO in different crystal structures. Obviously, going from Co to Ni corresponds to adding one electron to the system, namely $d^7 \rightarrow d^8$. According to the notion of crystal field theory, the available states of the metal-ion $d$-electrons split into submanifolds of different symmetry character when experiencing an electrostatic field from the ligand ions. In the octahedral oxygen environment of Co or Ni ions in the RS structure (point group $O_h$), the $t_{2g}$ states are lower in energy than the $e_g$ states, while in the tetrahedral ligand coordination in the ZB structure (point group $T_d$), the $t_2$ and $e$ manifolds exhibit the reversed energetic hierarchy. The orbitals carrying an index $g$ transform evenly (i.e., they do not change sign) under inversion. Since a tetrahedron has no inversion center, this symmetry operation is missing in $T_d$, and the subscript $g$ is dropped for the corresponding orbitals in the ZB structure [72]. With the $t_{2(g)}$ and $e_{(g)}$ levels being three- and two-fold degenerate, respectively, going from RS to ZB changes the filling pattern of electrons in those states. To analyze the electronic structure and the manifestation of crystal field theory, we calculated the spectral functions of NiO and CoO in RS and ZB structures by DFT+sicDMFT. The results are shown in Figure 4. The plots contain the total and O-projected spectral functions together with the local impurity spectra, which are further split into the $t_{2(g)}$ and $e_{(g)}$ contributions. As it follows from the fundamental properties of a spectral function, integrating it up to the Fermi level yields the number of electrons in the corresponding orbitals. We performed such integrations for the different contributions and provide the numbers in parentheses in the plots.

Starting with RS-NiO, we can clearly identify an occupation close to $t_{2g}^6 e_g^2$. This is well known [73], [74], [75] and follows the crystal field theory. In the occupied region, the energetic ordering is reflected by the half-occupied $e_g$ states lying closer to the Fermi edge than the fully occupied $t_{2g}$ states. Molecular orbital theory, which additionally considers the symmetry properties of the ligand ions, characterizes $t_{2g}$ as non-bonding and $e_g$ as $\sigma$-antibonding orbitals at a higher energy [76].

For RS-CoO, we find approximately $t_{2g}^5 e_g^2$, which is the widely reported configuration [65], [77], [78], [79]; it follows Hund's occupation rule with a resulting high-spin state. The low-spin state $t_{2g}^6 e_g^1$ is less favorable, since the crystal field-splitting is much smaller than the electron-electron repulsion within one orbital. Although less pronounced as in RS-NiO, the energy hierarchy between $t_{2g}$ and $e_g$ can still be identified in the occupied region and emerges clearly for the empty states, where the positions of the corresponding peak heights are shifted by about 0.6 eV with respect to each other. Experimental values of 1.1 eV from optical absorption [80] measurements and of 0.7 eV [19] are reported for the crystal-field splitting of RS-CoO. Setting this value to 0.5 eV in calculations based on a single-impurity Anderson model, Magnuson and coworkers obtained spectral functions in good agreement to measured data [64].

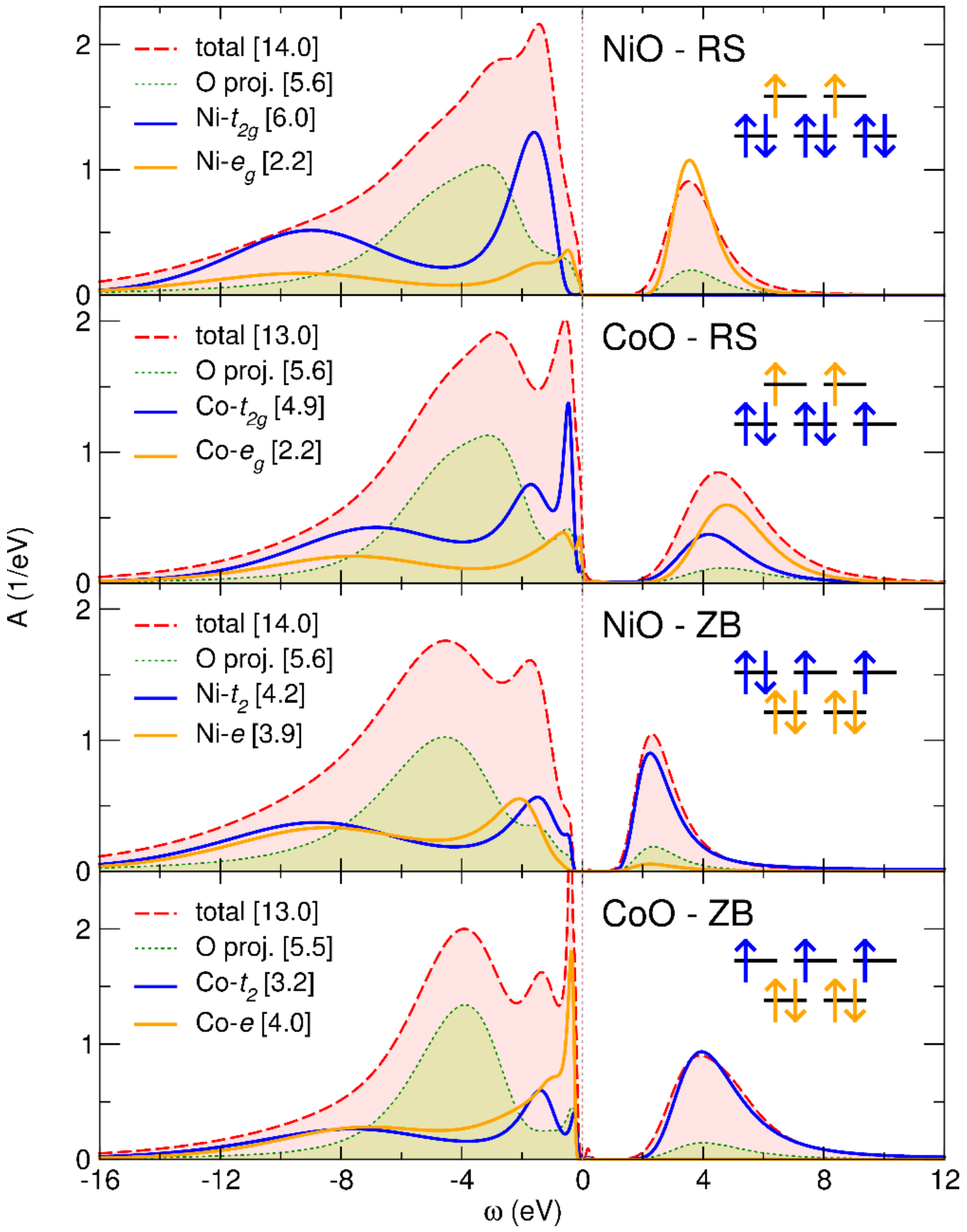

*Figure 4. Total, O-projected, and local $t_{2(g)}$ and $e_{(g)}$ spectral functions derived from DFT+sicDMFT for NiO and CoO in RS and ZB structures. For NiO, U = 10 eV and J = 1 eV, and for CoO, U = 9 eV and J = 1 eV were used. The numbers in parentheses are the values of the integrals of the respective curves up to the Fermi level ($\omega$ = 0 eV) and denote the occupation numbers. The level diagrams in the upper right do not reflect the true energy differences and levels but schematically relate the occupations to the expectations from crystal field theory. Arrows pointing up and down indicate electrons in different spin states, with colors corresponding to the $t_{2(g)}$ and $e_{(g)}$ lines.*

For the ZB structures, rounding the occupations derived from the integrated local spectral functions yields $e^4 t_2^4$ and $e^4 t_2^3$ for NiO and CoO, respectively, in line with the energy hierarchy expected from crystal field theory. Comparing RS-NiO to ZB-NiO, the shapes of the $t_{2g}$ and $e$ spectra qualitatively resemble each other, as well as those of the $e_g$ and $t_2$ curves, except for the spectral weights, which adjust according to the numbers of electrons. A similar observation holds for the two CoO structures, as, e.g., by the sharp peaks appearing in both, $t_{2g}$ (RS) and $e$ (ZB) spectra close to their valence band edges with smaller satellites at lower energies, as well as by the more balanced double peak structures of $e_g$ and $t_2$ character in both phases. Note that the absolute heights of the sharp peaks can vary with the settings of the analytical continuation, namely the noise-level parameter. However, the peak positions are largely independent of this choice.

Note that in the tetrahedral ligand environment of ZB, molecular orbital theory identifies $e$ as non-bonding and $t_2$ as antibonding electronic states. The experimentally observed most stable structure of paramagnetic NiO and CoO is RS, with each having two electrons in the anti-bonding $e_g$ states. The ZB structures are destabilized more strongly by 3 and 4 non-bonding electrons in the $t_2$ states of CoO and NiO, respectively, which explains why ZB-CoO has also been found in the energetically less favorable ZB structure [81], [82], whereas stoichiometric bulk NiO does not crystallize in this structure. One recent study, however, reports a ZB structure for thin films of $NiO_{1.2}$ [83].

It is apparent that when going from NiO to CoO, the total, O-projected, and local spectral functions exhibit sharper peaks and increased spectral weight in the low energy region below the Fermi level, independent of the crystal structure. For RS-NiO, the spectral substructure close to the valence band edge is discussed to result from two screening channels after electron removal, namely a local process corresponding to the final state $d^8L$ (similar to $d^7L$ described in Section 3.1 for RS-CoO), and a nonlocal screening leading to $d^8Z$ at a lower energy [32], [84], [85]. $Z$ denotes a hole in the strongly correlated Zhang-Rice state [86], which is formed by the hybridization of O($2p$) and Ni($3d$) states, and the $d^8Z$ entity arises from the interaction of holes in the respective orbitals. Increasing the number of holes in the valence configuration by partially replacing Ni in NiO with Li was shown earlier by authors of this work to lead to a considerable enhancement of the $d^8Z$ lowest energy peak [32]. Compared to NiO, CoO has one electron less in the $d$ states. In analogy to the Li-doped NiO, the effect of this additional hole on the CoO spectra is also seen in the height increase of the peak closest to the valence band edge, which can therefore be assumed to be of $d^7Z$ nature. The $d^7L$ channels of CoO then are manifested in the features at $-1.7$ eV (RS) and $-1.5$ eV (ZB), the former being visible only in the local spectral function and partly superimposed in the total spectrum by the O($2p$) main peak. Since the maximum of the latter is shifted by about $-1$ eV between RS and ZB, the $d^7L$ peak also appears in the total spectral function of ZB-CoO.

The spectral functions discussed above are obtained by analytical continuation of band- and orbital-projected Green's functions, which are related to the dynamical mean field and the electronic many-body self-energy by the Dyson equation [22]. Analyzing the converged self-energy $\Sigma$ yields insight into the nature and strength of correlation effects. We display in Figure 5 the imaginary parts of the self-energies which belong to the orbital-projected spectral functions shown in Figure 4.

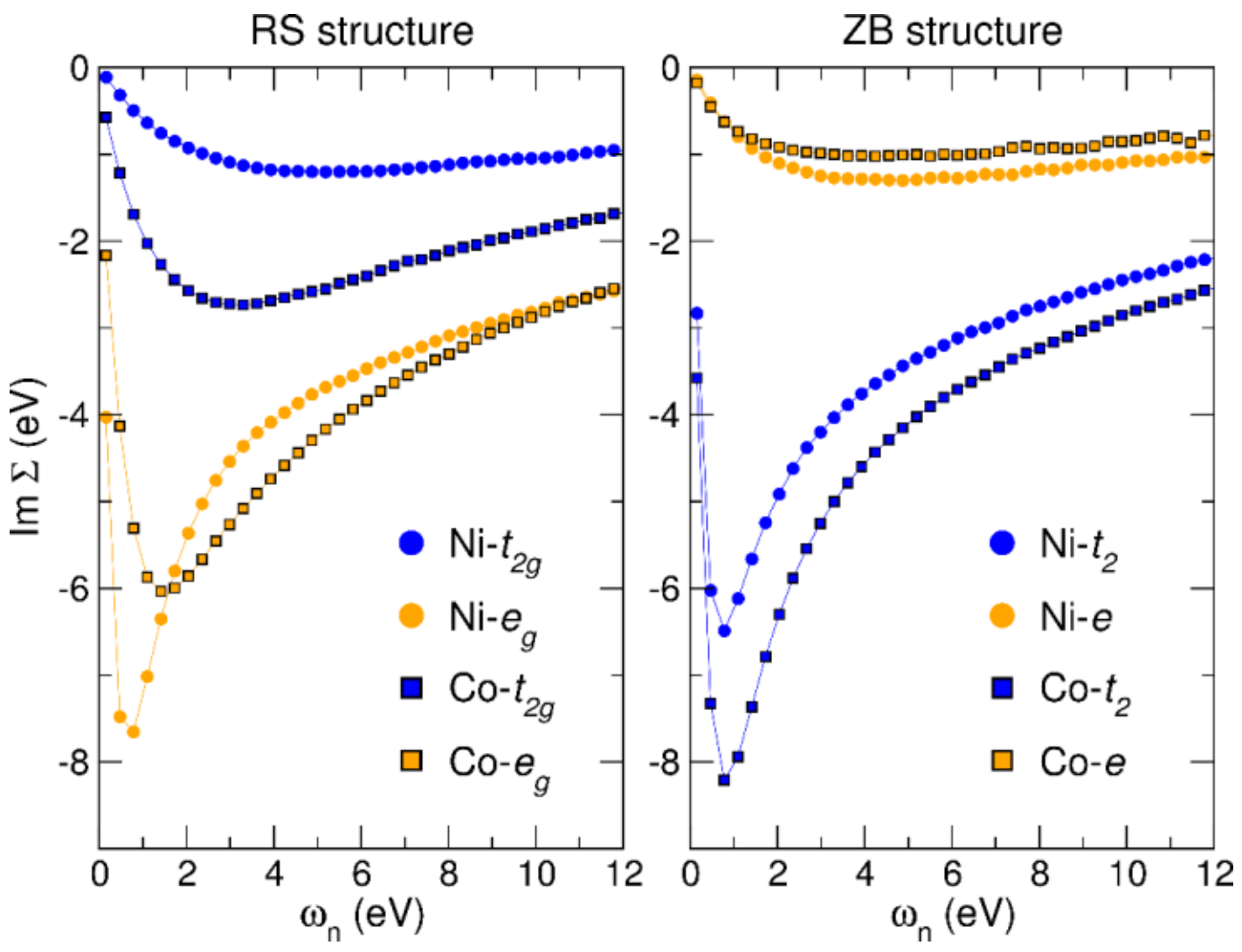

*Figure 5. Imaginary part $Im\ \Sigma(i\omega_n)$ of the self-energies of Ni- and Co-states of $t_{2(g)}$ and $e_{(g)}$ character for NiO and CoO in the RS (left panel) and in the ZB structure (right panel), as functions of the Matsubara frequency $\omega_n$. The lines are a guide for the eye. The coloring of the symbols is consistent with the coloring of the orbital-projected spectral functions and orbital occupation sketches of Figure 4.*

The lower (i.e., more negative) the imaginary part of the self-energy is at low Matsubara frequencies, the stronger the corresponding electrons are correlated. The fully occupied orbitals Ni-$t_{2g}$ in RS-NiO, Ni-$e$ in ZB-NiO, and Co-$e$ in ZB-CoO exhibit the weakest imaginary parts. The lack of one electron in the Co-$t_{2g}$ orbitals of RS-CoO causes correlation, which increases the absolute value of the respective self-energy. The largest negative self-energy values are present for the Ni-$e_g$ electrons in RS-NiO and for the Co-$t_2$ electrons in ZB-CoO. This corresponds in both cases to an exactly half-filled orbital manifold of those systems which contain besides this half-filled also a fully filled orbital manifold. Interestingly, the Co-$e_g$ orbitals in RS-CoO are also half-filled, but the additionally missing electron in the other orbital group of that system (Co-$t_{2g}$) results in a lower absolute value of the Co-$e_g$ self-energy at low frequencies and a higher absolute value at higher frequencies as compared to the situation where that other orbital group is completely filled (Ni-$e_g$/Ni-$t_{2g}$ in RS-NiO). Following the systems examined in Figure 4 from bottom to top, namely ZB-CoO, ZB-NiO, RS-CoO, and RS-NiO, the number of electrons in the $t_{2(g)}$ orbitals increases from 3 (half-filling), to 4, 5, and 6 (fully occupied), respectively. This hierarchy is exactly matched in the corresponding imaginary self-energies (all blue colored curves in Figure 5), having the highest to lowest absolute values, which is directly related to the combinatorial possibilities of electron arrangement in those states.

### 3.3. Influence of SIC and $U$ on the spectral functions

In this work, we calculated spectral functions for different compositions and structures by applying different settings for incorporating electron correlation in TM- and O-orbitals. Detailed comparisons of subsets of those curves were presented in Sections 3.1 and 3.2. To get an overview over the complete set of results, we extracted characteristic features from the TM- and O-projected spectral functions, namely the centers of gravity (centroids) of their occupied and unoccupied parts. The data are presented in Figure 6. The centroids indicate how spectral weight is shifted for the different settings.

At first, it is clearly seen that applying O($2p$)-SIC consistently moves the unoccupied TM-projected bands, which are essentially the upper Hubbard bands to higher values. In the occupied region, the corresponding tendencies are similar, but not as consistent. Instead of the centroid, which comprises the whole spectrum, the lower Hubbard band is expected to follow a more systematic behavior, namely a shift to higher energies, upon applying O($2p$)-SIC, too. The difference between upper and lower Hubbard band is therefore not strongly affected by SIC. The lower Hubbard band, however, cannot in all the systems be uniquely identified and separated from the background in our data. As explained in Section 3.1, the effective Coulomb penalty introduced by SIC keeps the upper Hubbard band and the occupied O($2p$) band at a larger distance $\Delta E_{\mathrm{pd}}$ by reducing the erroneous too strong charge fluctuations between the corresponding states. This behavior is apparent in Figure 6(b), where the $\Delta E_{\mathrm{pd}}$ values are given for all systems.

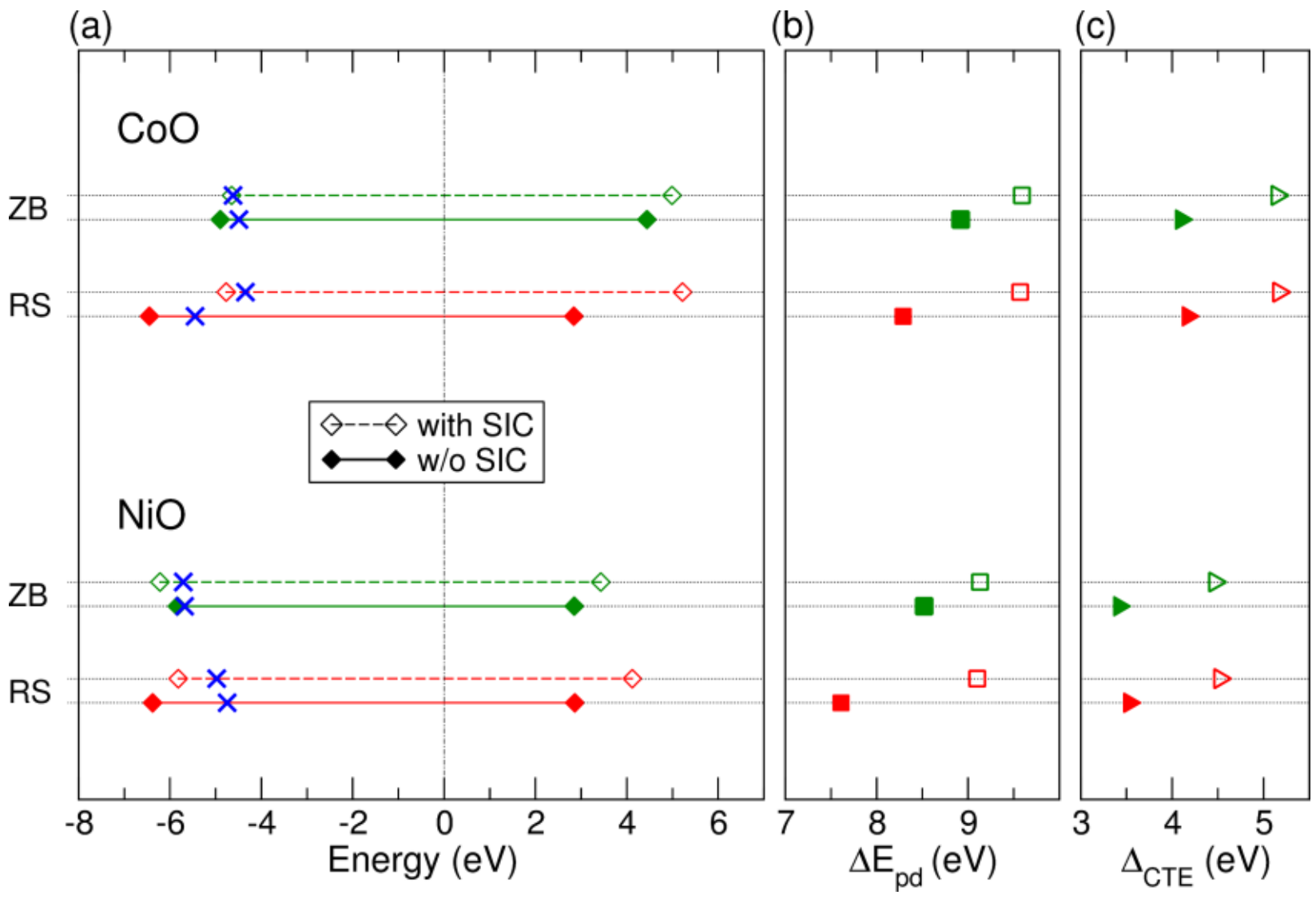

*Figure 6. Overview over energy centers of gravity (centroids) of element projected spectral functions for NiO (lower part) and CoO (upper part) in RS and ZB structures (denoted at the left vertical axis) calculated with DFT+(sic)DMFT [$U$(Ni-3$d$) = 10 eV, $U$(Co-3$d$) = 9 eV], with (dashed lines) and without ("w/o", solid lines) considering O(2p)-SIC. (a) diamond symbols correspond to the TM-projected spectral functions with the centroid of the occupied states on the left (negative energy) and the centroid of the unoccupied states on the right (positive energy). They are connected by lines for better visibility. Blue crosses correspond to the centroids of the occupied O-projected spectral functions. See the vertical arrows in Figure 3(c) for the positions of those centroids for RS-CoO, calculated with $U$(Co-3$d$) = 9 eV and with O(2p)-SIC. (b) differences $\Delta E_{pd}$ between unoccupied TM-projected [diamond symbols for positive energies in (a)] and occupied O-projected centroids [blue crosses in (a)]. (c) charge-transfer energy ($\Delta_{CTE}$) calculated from DFT and SIC-DFT.*

That diagram illustrates the subtle interplay between composition, structure, and the consideration of ligand correlation. If SIC is *not* applied, the gap-reducing charge fluctuations between the O($2p$) and the unoccupied TM($3d$) states act stronger in the RS than in the ZB phases, consistently for NiO and CoO. This is enabled by the six (RS) versus four (ZB) oxygen ligands of the TMs. If SIC is applied and the artificial gap-lowering effect is cured, the differences between $\Delta E_{\mathrm{pd}}$(RS) and $\Delta E_{\mathrm{pd}}$(ZB) vanish. Note that, looking only at the SIC-induced energy shifts of the unoccupied TM-$d$ peaks (positions of the diamonds in Figure 6(a) for positive energies), the largest shift is found for RS-CoO. That phase is the only one considered where unoccupied states of both, $t_{2(g)}$ and $e_{(g)}$ character, are involved in the charge fluctuations with the O($2p$) states, thereby being affected the most by applying SIC. The behavior of $\Delta E_{\mathrm{pd}}$ for the different structures and SIC treatments exhibit qualitative similarity in the NiO and CoO systems. Quantitatively, the values for CoO are shifted to higher values of about 0.5 eV on average with respect to those of NiO.

Note that the quantity $\Delta E_{\mathrm{pd}}$ is to be distinguished from what is commonly referred to as the *charge-transfer energy* $\Delta_{\mathrm{CTE}}$. The latter is an effective single-particle property, namely the energy required to transfer an electronic charge from an occupied O($2p$) to an unoccupied TM($3d$) state [15]. $\Delta_{\mathrm{CTE}}$ can therefore be calculated from averaged positions of single-particle energy levels $\varepsilon_d \equiv \varepsilon_{\mathrm{TM}(3d)}$ and $\varepsilon_p \equiv \varepsilon_{\mathrm{O}(2p)}$ ($\Delta_{\mathrm{CTE}} = \varepsilon_d - \varepsilon_p$), as, e.g., resulting from the projected densities of states of a DFT calculation (*cf.* Figure 1), rather than from the fully correlated spectral function of a DFT+sicDMFT calculation. Following Ref. [32], we calculated $\Delta_{\mathrm{CTE}}$ accordingly for the NiO and CoO structures with LDA and LDA+SIC. The results are shown in Figure 6(c). All systems exhibit a consistent shift of $\Delta_{\mathrm{CTE}}$ of about 1 eV upon applying SIC. Note that there are no band gaps in the so-derived densities of states since SIC was applied merely on the O($2p$) states in those calculations. However, the effect of SIC of separating the energies of the metal and ligand states ($pd$ splitting) is already present on the pure DFT level where no correlation is considered on the TM($3d$) states. In contrast to the $\Delta E_{\mathrm{pd}}$ values discussed above, which result from the fully correlated DFT+sicDMFT treatment, there is no difference between $\Delta_{\mathrm{CTE}}$ for RS and ZB structures. For RS structures, experimental values of the charge-transfer energy of 5.4 eV (NiO) and 6.1 eV (CoO) are reported [69], exhibiting a similar hierarchy and energy difference as we found for both, $\Delta E_{\mathrm{pd}}$ and $\Delta_{\mathrm{CTE}}$ values between the respective NiO and CoO systems. This is in line with the scheme of Zaanen, Sawatzki, and Allen, stating that the charge-transfer energy decreases for TMOs with TMs from left to right in the $3d$ row of the periodic table [15], defining the "early" TMOs as insulators of Mott-Hubbard type [$U$(TM-$3d$) $< \Delta_{\mathrm{CTE}}$] and the "late" TMOs (such as CoO and NiO) as insulators of charge-transfer type [$\Delta_{\mathrm{CTE}} < U$(TM-$3d$)].

In the common Hubbard model, the value of $U$(TM-$3d$) defines the distance between the upper and the lower Hubbard bands. Taking instead the distances between the centroids of the unoccupied and occupied TM-projected spectra, as derived for CoO, we obtain 7.1 eV for $U$(Co-$3d$) = 6 eV and 9.6 eV for $U$(Co-$3d$) = 9 eV when, in each case, averaged over RS and ZB structures, which is proportionally in line with the assumption. The results for each CoO system calculated in this study by DFT+(sic)DMFT are presented in Figure 7. We can identify different and combined effects of $U$(Co-$3d$) and O($2p$)-SIC on the spectral functions. On the one hand, as stated above, a larger value of $U$ consistently leads to larger differences between the centroids of occupied and unoccupied states, independent of the structure, and independent of whether SIC was applied or not. This effect can therefore be clearly attributed to $U$. On the other hand, almost consistently among all CoO systems, an effect which can be distinctively related to SIC manifests in a shift of both occupied and unoccupied centroid values to higher energies by a similar amount. The magnitude of this effect is larger for RS than for ZB structures. We can denote those two effects of $U$ and SIC as being of "first order" and independent of each other. An interrelated second order effect is also discernable from the results: for each of the 4 CoO systems defined by the structure and chosen $U$, the difference between unoccupied and occupied centroids, which is defined by $U$, consistently increases by 0.3 – 0.7 eV if SIC is applied.

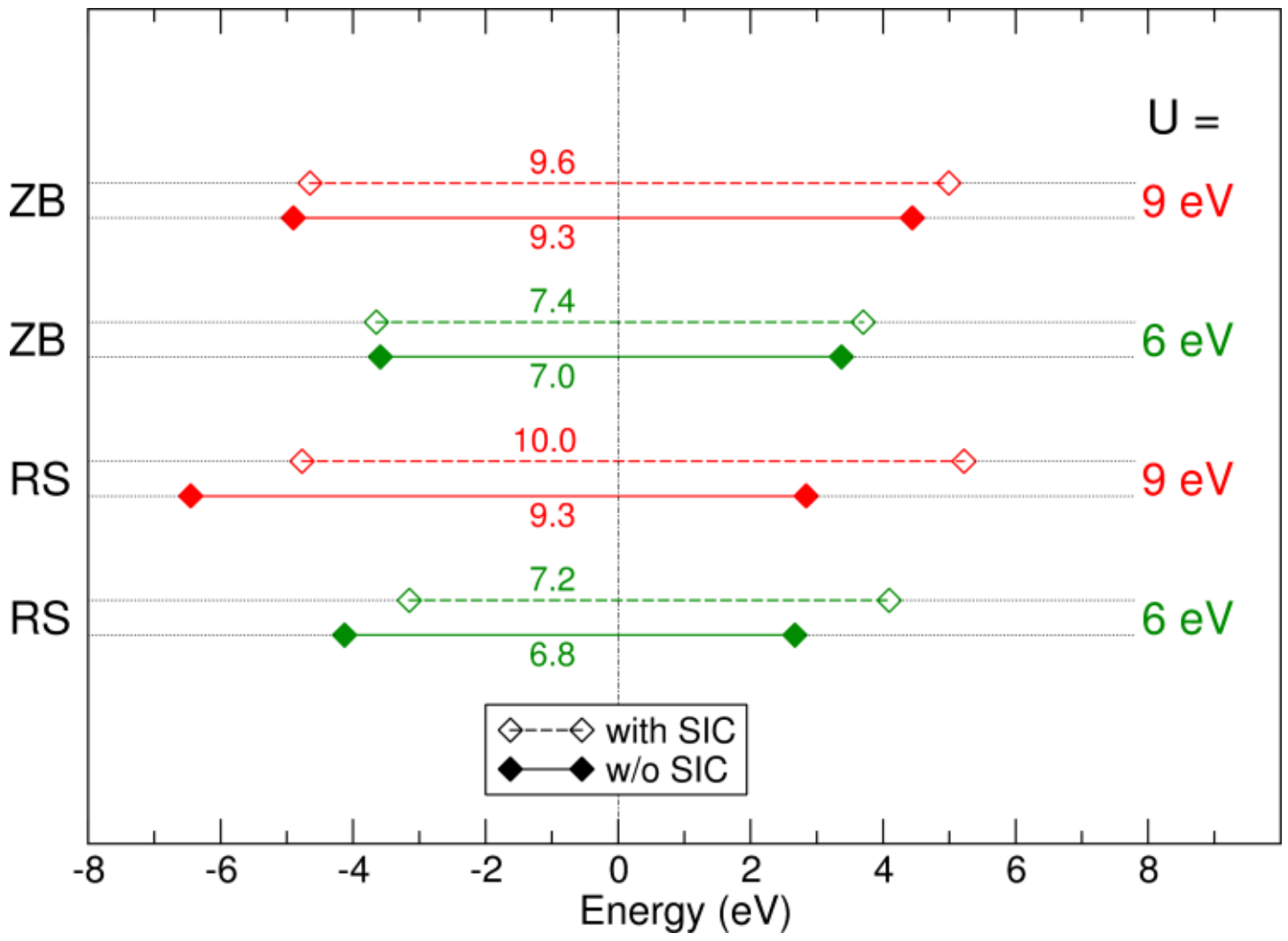

*Figure 7. Overview over energy centroids of Co-projected spectral functions for CoO in RS and ZB structures (denoted at the left vertical axis) calculated with DFT+(sic)DMFT for two different values of $U$(Co-$3d$), namely 6 eV and 9 eV. Data obtained with (dashed lines) and without ("w/o", solid lines) considering O($2p$)-SIC is displayed. Diamond symbols correspond to the centroids of occupied (negative energy) and unoccupied states (positive energy). See the vertical arrows in Figure 3(c) for the positions of those centroids for RS-CoO. For each system, the centroids are connected by horizontal lines for better visibility. The numbers next to the lines are the energy differences (lengths of the corresponding lines) in eV between the centroids of occupied and unoccupied states.*

## 4. Summary and conclusions

Characteristic features of bulk electron excitation spectra of correlated TMOs can serve as descriptors for the catalytic activity of those materials. In line with our previous work on RS-NiO, we demonstrated that also for RS-CoO an improvement in the spectral function, compared to experimental data, can be achieved upon combining DFT+DMFT with the SIC method (DFT+sicDMFT) to include both, correlation in the Co($3d$) and in the O($2p$) orbitals. This is mainly reflected in the distance of occupied and unoccupied bands of 5.1 eV close to the experimental value range (5 – 6 eV). The methodology was further applied to both NiO and CoO in the ZB structure to systematically analyze the effect of composition and ligand environment on the spectra. Expectations from crystal field theory concerning the occupation of the $3d$-submanifolds were confirmed, thereby assuming the lower stability of the ZB phases being caused by a larger number of electrons in antibonding states. For both compounds, NiO and CoO, the local (orbital resolved) spectral functions of the RS and ZB structures exhibit qualitative similarities when comparing the respective energetically lower and higher manifolds with each other, namely RS-$t_{2g}$ with ZB-$e$, and RS-$t_2$ with ZB-$e_g$. A possible explanation of a sharp low-energy peak at the valence band edge of the CoO structures is given by the increased incidence of Zhang-Rice states due to the reduced number of valence electrons compared to NiO, where Zhang-Rice behavior is widely discussed. We finally presented a systematic overview of the systems and correlation-related settings applied in this study with respect to centroids of oxygen and metal bands. This clearly revealed the effect of SIC, shifting the unoccupied bands and the energy difference to the occupied oxygen bands to higher energies. The described approach and presented results can support the research of highly correlated TMOs aiming for optimizing their functionality in energy conversion devices.

## ACKNOWLEDGMENTS

This work has been funded by the Ministry of Economic Affairs, Labour and Tourism Baden Württemberg through the Competence Center Quantum Computing Baden-Württemberg (KQCBW), projects QuESt and QuESt+, and by the Federal Ministry of Research, Technology and Space Germany (BMFTR) through the project QUBE. The authors gratefully acknowledge the Gauss Centre for Supercomputing e.V. (www.gauss-centre.eu) for funding this project by providing computing time through the John von Neumann Institute for Computing (NIC) on the GCS Supercomputer JUWELS at Jülich Supercomputing Centre (JSC). Structure figures were generated with VESTA [87].

**References**

[1] M. Imada, A. Fujimori, and Y. Tokura, “Metal-insulator transitions,” *Rev. Mod. Phys.* 70, 1039 (1998). doi: 10.1103/RevModPhys.70.1039

[2] Y. Tokura and N. Nagaosa, “Orbital physics in transition-metal oxides,” *Science* 288, 462, (2000). doi: 10.1126/science.288.5465.462.

[3] S.-W. Cheong, “The exciting world of orbitals,” *Nat. Mater.*, vol. 6, p. 927, 2007, doi: 10.1126/science.1149338.

[4] C. Ahn, A. Cavalleri, A. Georges, S. Ismail-Beigi, A. J. Millis, and J. M. Triscone, “Designing and controlling the properties of transition metal oxide quantum materials,” *Nat. Mater.*, vol. 20, p. 1462, 2021, doi: 10.1038/s41563-021-00989-2.

[5] D. S. Negi, D. Singh, P. A. Van Aken, and R. Ahuja, “Spin-entropy induced thermopower and spin-blockade effect in CoO,” *Phys. Rev. B*, vol. 100, p. 144108, 2019, doi: 10.1103/PhysRevB.100.144108.

[6] M. Taeño, D. Maestre, and A. Cremades, “An approach to emerging optical and optoelectronic applications based on NiO micro- And nanostructures,” *Nanophot.*, vol. 10, p. 1785, 2021, doi: 10.1515/nanoph-2021-0041.

[7] Y. Liu, C. Gao, Q. Li, and H. Pang, “Nickel Oxide/Graphene Composites: Synthesis and Applications,” *Chem. Eur. J.*, vol. 25, p. 2141, 2019, doi: 10.1002/chem.201803982.

[8] X. Zhan, Z. Wang, F. Wang, Z. Cheng, K. Xu, Q. Wang, M. Safdar, and J. He, “Efficient CoO nanowire array photocatalysts for H2 generation,” *Appl. Phys. Lett.*, vol. 105, p. 153903, 2014, doi: 10.1063/1.4898681.

[9] S. N. F. Moridon, M. I. Salehmin, M. A. Mohamed, K. Arifin, L. J. Minggu, and M. B. Kassim, “Cobalt oxide as photocatalyst for water splitting: Temperature-dependent phase structures,” *Int. J. Hydrog. Energ.*, vol. 44, p. 25495, 2019, doi: 10.1016/j.ijhydene.2019.08.075.

[10] H. Jung, A. Ma, S. A. Abbas, H. Y. Kim, H. R. Choe, S. Y. Jo, and K. M. Kim, “A new synthetic approach to cobalt oxides: Designed phase transformation for electrochemical water splitting,” *Chem. Eng. J.*, vol. 415, p. 127958, 2021, doi: 10.1016/j.cej.2020.127958.

[11] Z. Li, G. Luo, Z. Hu, X. Pei, Z. Zeng, S. Guo, J. Lv, S. Huang, Y. Wang, and X. Ma, “Zinc Blende CoO as an Efficient CO Nondissociative Adsorption Site for Direct Synthesis of Higher Alcohols from Syngas,” *ACS Catal.*, vol. 14, p. 2181, 2024, doi: 10.1021/acscatal.3c05579.

[12] W. T. Hong, K. A. Stoerzinger, Y.-L. Lee, L. Giordano, A. Grimaud, A. M. Johnson, J. Hwang, E. J. Crumlin, W. Yang, and Y. Shao-Horn, “Charge-transfer-energy-dependent oxygen evolution reaction mechanisms for perovskite oxides,” *Energy Environ. Sci.*, vol. 10, p. 2190, 2017, doi: 10.1039/c7ee02052j.

[13] G. Kotliar and D. Vollhardt, “Strongly Correlated Materials: Insights From Dynamical Mean-Field Theory,” *Phys. Today*, vol. 57, p. 53, 2004, doi: 10.1063/1.1712502.

[14] R. Zimmermann, P. Steiner, R. Claessen, F. Reinert, S. Hüfner, P. Blaha, and P. Dufek, “Electronic structure of 3d-transition-metal oxides: On-site Coulomb repulsion versus covalency,” *J. Phys. Condens. Matter*, vol. 11, p. 1657, 1999, doi: 10.1088/0953-8984/11/7/002.

[15] J. Zaanen, G. A. Sawatzky, and J. W. Allen, “Band gaps and electronic structure of transition-metal compounds,” *Phys. Rev. Lett.*, vol. 55, p. 418, 1985, doi: 10.1103/PhysRevLett.55.418.

[16] A. Fujimori, T. Yoshida, K. Okazaki, T. Tsujioka, K. Kobayashi, T. Mizokawa, M. Onoda, T. Katsufuji, Y. Taguchi, and Y. Tokura, “Electronic structure of Mott-Hubbard-type transition-metal oxides,” *J. Electron Spectrosc. Relat. Phenom.*, vol. 117–118, p. 277, 2001, doi: 10.1016/S0368-2048(01)00253-5.

[17] T. M. Schuler, D. L. Ederer, S. Itza-Ortiz, G. T. Woods, T. A. Callcott, and J. C. Woicik, “Character of the insulating state in NiO: A mixture of charge-transfer and Mott-Hubbard character,” *Phys. Rev. B*, vol. 71, p. 115113, 2005, doi: 10.1103/PhysRevB.71.115113.

[18] Z.-X. Shen, J. W. Allen, P. A. P. Lindberg, D. S. Dessau, B. O. Wells, A. Borg, W. Ellis, J. S. Kang, S.-J. Oh, I. Lindau, and W. E. Spicer, “Photoemission study of CoO,” *Phys. Rev. B*, vol. 42, p. 1817, 1990.

[19] J. van Elp, J. L. Wieland, H. Eskes, P. Kuiper, G. A. Sawatzky, F. M. F. de Groot, and T. S. Turner, “Electronic Structure of CoO, Li-doped CoO, and LiCoO2,” *Phys. Rev. B*, vol. 44, p. 6090, 1991.

[20] R. Sakuma and F. Aryasetiawan, “First-principles calculations of dynamical screened interactions for the transition metal oxides MO (M=Mn, Fe, Co, Ni),” *Phys. Rev. B*, vol. 87, p. 165118, 2013, doi: 10.1103/PhysRevB.87.165118.

[21] V. I. Anisimov, A. I. Poteryaev, M. A. Korotin, A. O. Anokhin, and G. Kotliar, “First-principles calculations of the electronic structure and spectra of strongly correlated systems: dynamical mean-field theory,” *J. Phys. Condens. Matter*, vol. 9, p. 7359, 1997, doi: 10.1088/0953-8984/9/4/002.

[22] F. Lechermann, A. Georges, A. Poteryaev, S. Biermann, M. Posternak, A. Yamasaki, and O. K. Andersen, “Dynamical mean-field theory using Wannier functions: A flexible route to electronic structure calculations of strongly correlated materials,” *Phys. Rev. B*, vol. 74, p. 125120, 2006, doi: 10.1103/PhysRevB.74.125120.

[23] J. Kuneš, V. I. Anisimov, A. V. Lukoyanov, and D. Vollhardt, “Local correlations and hole doping in NiO: A dynamical mean-field study,” *Phys. Rev. B*, vol. 75, p. 165115, 2007, doi: 10.1103/PhysRevB.75.165115.

[24] I. A. Nekrasov, N. S. Pavlov, and M. V. Sadovskii, “Consistent LDA’ + DMFT approach to the electronic structure of transition metal oxides: Charge transfer insulators and correlated metals,” *J. Exp. Theor. Phys.*, vol. 116, p. 620, 2013, doi: 10.1134/S1063776113030126.

[25] A. Hariki, T. Uozumi, and J. Kuneš, “LDA+DMFT approach to core-level spectroscopy: Application to 3d transition metal compounds,” *Phys. Rev. B*, vol. 96, p. 045111, 2017, doi: 10.1103/PhysRevB.96.045111.

[26] P. Wei and Z. Qing Qi, “Insulating gap in the transition-metal oxides: A calculation using the local-spin-density approximation with the on-site Coulomb U correlation correction,” *Phys. Rev. B*, vol. 49, p. 10864, 1994.

[27] A. S. Moskvin, “DFT, L(S)DA, LDA+U, LDA+DMFT, …, whether we do approach to a proper description of optical response for strongly correlated systems?,” *Opt. Spectrosc.*, vol. 121, p. 467, 2016, doi: 10.1134/S0030400X16100167.

[28] S. Mandal, K. Haule, K. M. Rabe, and D. Vanderbilt, “Systematic beyond-DFT study of binary transition metal oxides,” *npj Comp. Mater.*, vol. 5, p. 115, Nov. 2019, doi: 10.1038/s41524-019-0251-7.

[29] A. Georges, G. Kotliar, W. Krauth, and M. J. Rozenberg, “Dynamical mean-field theory of strongly correlated fermion systems and the limit of infinite dimensions,” *Rev. Mod. Phys.*, vol. 68, p. 13, 1996, doi: 10.1007/s12031-008-9042-1.

[30] D. Vogel, P. Krüger, and J. Pollmann, “Self-interaction and relaxation-corrected pseudopotentials for II-VI semiconductors,” *Phys. Rev. B*, vol. 54, p. 5495, 1996.

[31] W. Körner and C. Elsässer, “First-principles density functional study of dopant elements at grain boundaries in ZnO,” *Phys. Rev. B*, vol. 81, p. 085324, 2010, doi: 10.1103/PhysRevB.81.085324.

[32] F. Lechermann, W. Körner, D. F. Urban, and C. Elsässer, “Interplay of charge-transfer and Mott-Hubbard physics approached by an efficient combination of self-interaction correction and dynamical mean-field theory,” *Phys. Rev. B*, vol. 100, p. 115125, 2019, doi: 10.1103/PhysRevB.100.115125.

[33] F. Lechermann, “Late transition metal oxides with infinite-layer structure: Nickelates versus cuprates,” *Phys. Rev. B*, vol. 101, p. 081110(R), 2020, doi: 10.1103/PhysRevB.101.081110.

[34] F. Lechermann, “Assessing the correlated electronic structure of lanthanum nickelates,” *Electron. Struct.*, vol. 4, p. 15005, 2022, doi: 10.1088/2516-1075/ac5c6a.

[35] J. Gondolf, I. M. Eremin, and F. Lechermann, “Revisiting the metal-to-metal transition in 2H-AgNiO2,” *EPL*, vol. 144, p. 36003, 2023, doi: 10.1209/0295-5075/ad0d1d.

[36] A. Carta, A. Panda, and C. Ederer, “Importance of ligand on-site interactions for the description of Mott-insulators in DFT+DMFT,” *npj Comp. Mater.*, vol. 12, p. 57, 2026, doi: 10.1038/s41524-025-01928-4.

[37] G. Kotliar, S. Y. Savrasov, K. Haule, V. S. Oudovenko, O. Parcollet, and C. A. Marianetti, “Electronic structure calculations with dynamical mean-field theory,” *Rev. Mod. Phys.*, vol. 78, p. 865, 2006, doi: 10.1103/RevModPhys.78.865.

[38] Dm. Korotin, A. V. Kozhevnikov, S. L. Skornyakov, I. Leonov, N. Binggeli, V. I. Anisimov, and G. Trimarchi, “Construction and solution of a Wannier-functions based Hamiltonian in the pseudopotential plane-wave framework for strongly correlated materials,” *Eur. Phys. J. B*, vol. 65, p. 91, 2008, doi: 10.1140/epjb/e2008-00326-3.

[39] V. I. Anisimov and A. V. Lukoyanov, “Investigation of real materials with strong electronic correlations by the LDA+DMFT method,” *Acta Cryst.*, vol. C70, p. 137, 2014, doi: 10.1107/S2053229613032312.

[40] B. Amadon, F. Lechermann, A. Georges, F. Jollet, T. O. Wehling, and A. I. Lichtenstein, “Plane-wave based electronic structure calculations for correlated materials using dynamical mean-field theory and projected local orbitals,” *Phys. Rev. B*, vol. 77, p. 205112, 2008, doi: 10.1103/PhysRevB.77.205112.

[41] C. Elsässer, N. Takeuchi, K. M. Ho, C. T. Chan, P. Braun, and M. Fähnle, “Relativistic

effects on ground state properties of 4d and 5d transition metals," *J. Phys. Condens. Matter*, vol. 2, p. 4371, 1990, doi: 10.1088/0953-8984/2/19/006.

[42] B. Meyer, C. Elsässer, F. Lechermann, and M. Fähnle, "FORTRAN 90 Program for Mixed-Basis-Pseudopotential Calculations for Crystals," *Max-Planck-Institut für Metallforschung, Stuttgart (unpublished)*.

[43] H. Park, A. J. Millis, and C. A. Marianetti, "Computing total energies in complex materials using charge self-consistent DFT+DMFT," *Phys. Rev. B*, vol. 90, p. 235103, 2014, doi: 10.1103/PhysRevB.90.235103.

[44] S. Bhandary and K. Held, "Self-energy self-consistent density functional theory plus dynamical mean field theory," *Phys. Rev. B*, vol. 103, p. 245116, 2021, doi: 10.1103/PhysRevB.103.245116.

[45] S. Beck, A. Hampel, O. Parcollet, C. Ederer, and A. Georges, "Charge self-consistent electronic structure calculations with dynamical mean-field theory using Quantum ESPRESSO, Wannier 90 and TRIQS," *J. Phys. Condens. Matter*, vol. 34, p. 235601, 2022, doi: 10.1088/1361-648X/ac5d1c.

[46] A. Valli, M. P. Bahlke, A. Kowalski, M. Karolak, C. Herrmann, and G. Sangiovanni, "Kondo screening in Co adatoms with full Coulomb interaction," *Phys. Rev. Res.*, vol. 2, p. 033432, 2020, doi: 10.1103/PhysRevResearch.2.033432.

[47] V. I. Anisimov, I. V. Solovyev, M. A. Korotin, M. T. Czyzyk, and G. A. Sawatzky, "Density-functional theory and NiO photoemission spectra," *Phys. Rev. B*, vol. 48, p. 16929, 1993, doi: 10.1103/PhysRevB.48.16929.

[48] A. I. Liechtenstein, V. I. Anisimov, and J. Zaanen, "Density-functional theory and strong interactions: Orbital ordering in Mott-Hubbard insulators," *Phys. Rev. B*, vol. 52, p. R5467, 1995.

[49] I. Leonov, L. Pourovskii, A. Georges, and I. A. Abrikosov, "Magnetic collapse and the behavior of transition metal oxides at high pressure," *Phys. Rev. B*, vol. 94, p. 155135, 2016, doi: 10.1103/PhysRevB.94.155135.

[50] S. K. Panda, H. Jiang, and S. Biermann, "Pressure dependence of dynamically screened Coulomb interactions in NiO: Effective Hubbard, Hund, intershell, and intersite components," *Phys. Rev. B*, vol. 96, p. 045137, 2017, doi: 10.1103/PhysRevB.96.045137.

[51] L. Zhang, P. Staar, A. Kozhevnikov, Y.-P. Wang, J. Trinastic, T. Schulthess, and H.-P. Cheng, "DFT+DMFT calculations of the complex band and tunneling behavior for the transition metal monoxides MnO, FeO, CoO, and NiO," *Phys. Rev. B*, vol. 100, p. 035104, 2019, doi: 10.1103/PhysRevB.100.035104.

[52] W. Körner and C. Elsässer, "Density functional theory study of dopants in polycrystalline TiO2," *Phys. Rev. B*, vol. 83, p. 205315, 2011, doi: 10.1103/PhysRevB.83.205315.

[53] W. Körner, P. Gumbsch, and C. Elsässer, "Analysis of electronic subgap states in amorphous semiconductor oxides based on the example of Zn-Sn-O systems," *Phys. Rev. B*, vol. 86, p. 165210, 2012, doi: 10.1103/PhysRevB.86.165210.

[54] D. F. Urban, W. Körner, and C. Elsässer, "Mechanisms for p -type behavior of ZnO, Zn1-xMgx O, and related oxide semiconductors," *Phys. Rev. B*, vol. 94, p. 075140, 2016, doi: 10.1103/PhysRevB.94.075140.

[55] P. Seth, I. Krivenko, M. Ferrero, and O. Parcollet, "TRIQS/CTHYB: A continuous-time quantum Monte Carlo hybridisation expansion solver for quantum impurity problems," *Comp. Phys. Commun.*, vol. 200, p. 274, 2016, doi: 10.1016/j.cpc.2015.10.023.

[56] M. Jarrell and J. E. Gubernatis, “Bayesian inference and the analytic continuation of imaginary-time quantum Monte Carlo data,” *Phys. Rep.*, vol. 269, p. 133, 1996, doi: 10.1016/0370-1573(95)00074-7.

[57] S. P. Patil, L. D. Jadhav, D. P. Dubal, and V. R. Puri, “Characterization of NiO-Al2O3 composite and its conductivity in biogas for solid oxide fuel cell,” *Mater. Sci.-Pol.*, vol. 34, p. 266, 2016, doi: 10.1515/msp-2016-0045.

[58] Y. C. Sui, Y. Zhao, J. Zhang, S. Jaswal, X. Z. Li, and D. J. Sellmyer, “Ferromagnetic multipods fabricated by solution phase synthesis and hydrogen reduction,” *IEEE Trans. Magn.*, vol. 43, p. 3115, 2007, doi: 10.1109/TMAG.2007.894204.

[59] H. Monkhorst and J. Pack, “Special points for Brillouin zone integrations,” *Phys. Rev. B*, vol. 13, p. 5188, 1976, doi: 10.1103/PhysRevB.13.5188.

[60] H. Park, R. Nanguneri, and A. T. Ngo, “DFT+DMFT study of spin-charge-lattice coupling in covalent LaCoO3,” *Phys. Rev. B*, vol. 101, p. 195125, 2020, doi: 10.1103/PhysRevB.101.195125.

[61] M. D. Rechtin and B. L. Averbach, “Tetragonal elongation in CoO near the Néel point,” *Phys. Rev. Lett.*, vol. 26, p. 1483, 1971, doi: 10.1103/PhysRevLett.26.1483.

[62] S.-P. Jeng and V. E. Henrich, “Direct Observation of Defect-Induced Surface Covalency in Ionic CoO (100),” *Solid State Commun.*, vol. 75, p. 1013, 1990.

[63] M. A. Langell, M. D. Anderson, G. A. Carson, L. Peng, and S. Smith, “Valence-band electronic structure of Co3O4 epitaxy on CoO(100),” *Phys. Rev. B*, vol. 59, p. 4791, 1999, doi: 10.3166/acsm.40.103-109.

[64] M. Magnuson, S. M. Butorin, J. H. Guo, and J. Nordgren, “Electronic structure investigation of CoO by means of soft x-ray scattering,” *Phys. Rev. B*, vol. 65, p. 205106, 2002, doi: 10.1103/PhysRevB.65.205106.

[65] H. A. E. Hagelin-Weaver, G. B. Hoflund, D. M. Minahan, and G. N. Salaita, “Electron energy loss spectroscopic investigation of Co metal, CoO, and Co3O4 before and after Ar+ bombardment,” *Appl. Surf. Sci.*, no. 235, p. 420, 2004, doi: 10.1016/j.apsusc.2004.02.062.

[66] I. Leonov, A. O. Shorikov, V. I. Anisimov, and I. A. Abrikosov, “Emergence of quantum critical charge and spin-state fluctuations near the pressure-induced Mott transition in MnO, FeO, CoO, and NiO,” *Phys. Rev. B*, vol. 101, p. 245144, 2020, doi: 10.1103/PhysRevB.101.245144.

[67] G. A. Sawatzky and J. W. Allen, “Magnitude and origin of the band gap in NiO,” *Phys. Rev. Lett.*, vol. 53, p. 2339, 1984, doi: 10.1103/PhysRevLett.53.2339.

[68] G. Lee and S. J. Oh, “Electronic structures of NiO, CoO, and FeO studied by 2p core-level x-ray photoelectron spectroscopy,” *Phys. Rev. B*, vol. 43, p. 14674, 1991, doi: 10.1103/PhysRevB.43.14674.

[69] E. Z. Kurmaev, R. G. Wilks, A. Moewes, L. D. Finkelstein, S. N. Shamin, and J. Kuneš, “Oxygen x-ray emission and absorption spectra as a probe of the electronic structure of strongly correlated oxides,” *Phys. Rev. B*, vol. 77, p. 165127, 2008, doi: 10.1103/PhysRevB.77.165127.

[70] A. A. Dyachenko, A. O. Shorikov, A. V. Lukoyanov, and V. I. Anisimov, “LDA+DMFT study of magnetic transition and metallization in CoO under pressure,” *JETP Lett.*, vol. 96, p. 56, 2012, doi: 10.1134/S002136401213005X.

[71] L. Huang, Y. Wang, and X. Dai, “Pressure-driven orbital selective insulator-to-metal transition and spin-state crossover in cubic CoO,” *Phys. Rev. B*, vol. 85, p. 245110,

2012, doi: 10.1103/PhysRevB.85.245110.

[72] B. N. Figgis and M. A. Hitchman, *Ligand Field Theory and its Applications*. Wiley-VCH, 2000.

[73] L. F. Mattheiss, “Electronic Structure of the 3d Transition-Matal Monoxides. II. Interpretation,” *Phys. Rev. B*, vol. 5, p. 306, 1972.

[74] A. Fujimori, F. Minami, and S. Sugano, “Multielectron satellites and spin polarization in photoemission from Ni compounds,” *Phys. Rev. B*, vol. 29, p. 5225, 1984, doi: 10.1103/PhysRevB.29.5225.

[75] S. Hüfner, “Electronic structure of NiO and related 3d-transition-metal compounds,” *Adv. Phys.*, vol. 43, p. 183, 1994, doi: 10.1080/00018739400101495.

[76] J. S. Griffith and L. E. Orgel, “Ligand-Field Theory,” *Q. Rev. Chem. Soc.*, vol. 11, p. 381, 1957, doi: 10.1039/QR9571100381.

[77] K. S. Kim, “X-ray-photoelectron spectroscopic studies of the electronic structure of CoO,” *Phys. Rev. B*, vol. 11, p. 2177, 1975, doi: 10.1103/PhysRevB.11.2177.

[78] S. Shi and V. Staemmler, “Ab initio study of local d-d excitations in bulk CoO, at the CoO(100) surface, and in octahedral Co2+ complexes,” *Phys. Rev. B*, vol. 52, p. 12345, 1995, doi: 10.1103/PhysRevB.52.12345.

[79] H. X. Deng, J. Li, S. S. Li, J. B. Xia, A. Walsh, and S. H. Wei, “Origin of antiferromagnetism in CoO: A density functional theory study,” *Appl. Phys. Lett.*, vol. 96, p. 162508, 2010, doi: 10.1063/1.3402772.

[80] G. W. Pratt Jr. and R. Coelho, “Optical Absorption of CoO and MnO above and below the Néel Temperature,” *Phys. Rev.*, vol. 116, p. 281, 1959.

[81] M. J. Redman and E. G. Steward, “Cobaltous oxide with the zinc blende/wurtzite-type crystal structure,” *Nature*, vol. 193, p. 867, 1962, doi: 10.1038/193867a0.

[82] J. Dicarlo and A. Navrotskyf, “Energetics of Cobalt(II) Oxide with the Zinc-Blende Structure,” *J. Am. Ceram. Soc.*, vol. 76, p. 2465, 1993.

[83] D. Fischer, “Nickel oxide films with the zinc blende-type structure – A re-evaluation of X-ray diffraction data,” *Mater. Today Comm.*, vol. 41, p. 110681, 2024, doi: 10.1016/j.mtcomm.2024.110681.

[84] M. Taguchi, M. Matsunami, Y. Ishida, R. Eguchi, A. Chainani, Y. Takata, M. Yabashi, K. Tamasaku, Y. Nishino, T. Ishikawa, Y. Senba, H. Ohashi, and S. Shin, “Revisiting the valence-band and core-level photoemission spectra of NiO,” *Phys. Rev. Lett.*, vol. 100, p. 206401, 2008, doi: 10.1103/PhysRevLett.100.206401.

[85] A. Hariki, Y. Ichinozuka, and T. Uozumi, “Dynamical Mean-Field Approach to Ni 2p X-ray Photoemission Spectra of NiO: A Role of Antiferromagnetic Ordering,” *J. Phys. Soc. Jpn.*, vol. 82, p. 043710, 2013, doi: 10.7566/JPSJ.82.043710.

[86] F. C. Zhang and T. M. Rice, “Effective Hamiltonian for the superconducting Cu oxides,” *Phys. Rev. B*, vol. 37, p. 3759, 1988, doi: 10.1103/PhysRevB.37.3759.

[87] K. Momma and F. Izumi, “VESTA 3 for three-dimensional visualization of crystal , volumetric and morphology data,” *J. Appl. Crystallogr.*, vol. 44, p. 1272, 2011, doi: 10.1107/S0021889811038970.